\documentclass[aps,prd,twocolumn,showpacs,superscriptaddress]{revtex4-2}
\usepackage{latexsym,amssymb,amsmath,graphicx}

\usepackage[unicode,colorlinks,pagebackref]{hyperref}
\makeatletter
\renewcommand{\@biblabel}[1]{#1.}
\makeatother

\begin{document}
\title{Gravitational dephasing in spontaneous emission of atomic ensembles in timed Dicke states}
\author{V.~Stefanov}
\affiliation{Institute of Physics of Belarus NAS, Nezavisimosti Ave. 68, Minsk 220072 Belarus}
\affiliation{ICRANet-Minsk of  Belarus NAS, Nezavisimosti Ave. 68, Minsk 220072 Belarus}
\author{I.~Siutsou}
\affiliation{ICRANet-Minsk of  Belarus NAS, Nezavisimosti Ave. 68, Minsk 220072 Belarus}
\author{D.~Mogilevtsev}
\affiliation{Institute of Physics of Belarus NAS, Nezavisimosti Ave. 68, Minsk 220072 Belarus}

\begin{abstract}
Here we discuss an effect of dephasing induced by weak gravitational field on the collective radiation dynamics of atomic system in timed single-photon Dicke states. We show that a photon absorbed by the stationary system of randomly placed stationary atoms is no more spontaneously emitted in the direction of the impinging photon. The influence of gravity leads to broadening of the angular distribution of emission . %Even for the spherically symmetric gravitational field, the broadening has specific non-symmetrical character.

\end{abstract}
%\PhySH{Open quantum systems & decoherence, Quantum fields in curved spacetime, Quantum state engineering}

\maketitle

\section{Introduction}

Nowadays, interplay of quantumness and gravity attracts a lot of attention \cite{Woodard}. Recently it was recognized that even weak gravitation can have rather noticeable effects on quantum interference and on quantum correlations, especially on those of large systems \cite{gambini, Blencowe, wang17v1, wangsci, wang17v2, zych2012, pik12, bassi16v1, bassi16v2, pik15, pik17, plenio17, diosi17, khalili, asen, hilweg, khosla17, manisc, margalit16, unruh14, unruh15}. These effects and their description are already well beyond the region of purely theoretical speculations. For example,  gravimeters based on gravitationally induced atom interference are very promising in stability and accuracy and are under active development now \cite{peters1999, enc2016, menoret2018}. 

It is already established that an influence of gravitational field on quantum interference might be quite destructive. Even in the linear approximation allowing to quantize in a standard way the gravitational field, quantum fluctuations of the gravity unavoidably lead to appearance of decoherence \cite{Blencowe, wang17v1, wangsci, wang17v2}. Another kind of decoherence arising due to interaction of the gravitational field with a quantum particle can be captured even when considering classical gravitational field. This is so-called "time-dilation decoherence" \cite{zych2012, pik12, bassi16v1, bassi16v2, pik15, pik17}. The essence of this effect can be described with the simple interferometric example \cite{pik15, margalit16}. If two interfering particles are moving through different arms of the interferometer, and these arms are subjected to different gravitational field interacting with inner degrees of freedom of the particles, the visibility of the interference measured at the output of the interferometer would be lessened. This observable effect does not depend of the frame, and can be seen even with photons. For an entangled state of a large number of particles, this decoherence can be noticeable even near the Earth surface \cite{bassi16v1, bassi16v2}.

Here we discuss another effects stemming from interaction of a set of entangled quantum systems with classical gravity. We consider how the gravity affects creation and spontaneous decay of single timed Dicke state of an extended atomic system. Emission of a photon previously absorbed by a set of randomly placed two-level atoms demonstrates quite curious and counterintuitive effect: strictly directional emission. If a photon is absorbed by sufficiently large system of identical non-interacting two-level atoms, a collective entangled state (so called "timed Dicke state") is formed. Then, this state leads to the spontaneous emission of a photon precisely in the direction of the photon that was absorbed \cite{scully2006, boag2013}. Noticeably, the effect does not depend on positions of each particular atom. The prerequisite is to have the sum of phase factors stemming from different atomic positions tending to the delta-function (which is provided by random placement of sufficiently large number of atoms).

In this work we show how this effect of directional emission is broken by gravity. Curiously, it appears that basic features of the effect are retained in the presence of gravity: a pure entangled "gravity-affected timed Dicke state" is formed after photon absorption, emitted field does not depend on the size of the atomic system. However, different time-dilation in different parts of the system leads to the broadening of the angular distribution of the emitted photon. This broadening is asymmetric with respect to the direction of the gravity gradient.

The outline of this paper is as follows. In Section II we describe the process of timed Dicke states formation and spontaneous directional emission in flat time-space. Then, in Section III we consider gravitational field produced by spherically symmetric mass distribution implementing Schwarzschild metrics and introduce corrections to electromagnetic field eigenmodes produced by gravity. In Section IV we derive the effective Hamiltonian describing atom-field interaction leading to the spontaneous emission and consider the emission in the Markovian approximation. Finally, in Section V and VI we analyse gravitational corrections to the directional emission effect and derive the angular distribution of the emitted single-photon field. Conclusions follow.

\begin{figure}[htb]
\includegraphics[width=\linewidth]{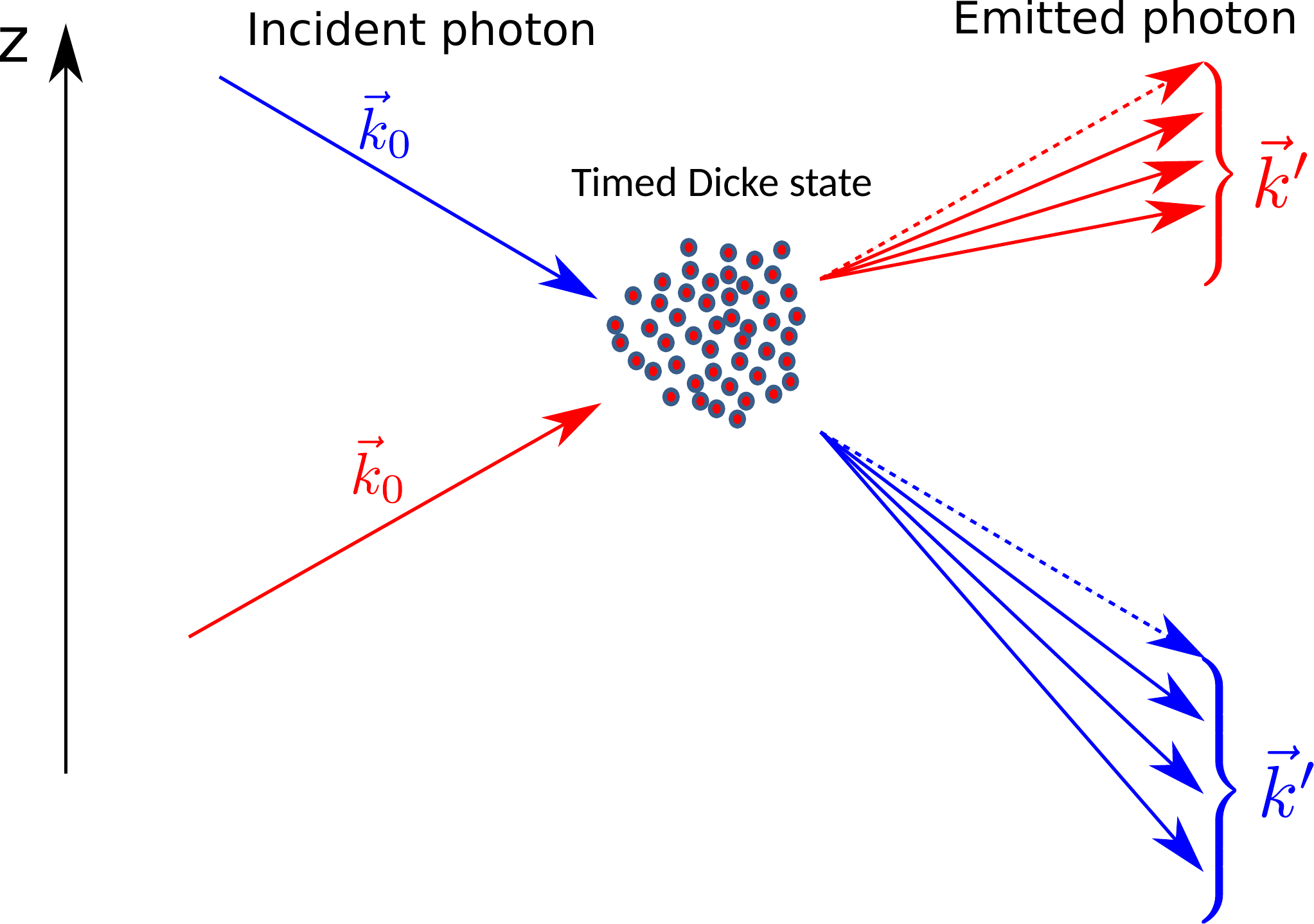}
\caption{A schematic depiction of the timed Dicke states creation by the absorbed photon in the emitters cloud and consequent emission.}
\label{fig1}
\end{figure}

\section{Timed Dicke state in flat time-space}
\label{sec1}
We start considering the system of $N$ stationary identical two-level atoms (TLA) interacting in the dipole-dipole and rotating-wave approximations with the modes of the electromagnetic field in homogeneous vacuum \cite{scully2006}. Such a system is described by the following standard Hamiltonian
\begin{eqnarray}
\nonumber
 \hat{H}=\sum\limits_{\textbf{k}} \hbar \omega_{\textbf{k}} \hat{q}_{\textbf{k}}^\dagger \hat{q}_{\textbf{k}}+\frac{1}{2}\sum\limits_j \hbar \nu\hat{(\sigma_z)}_j+\\
 \sum\limits_{{\textbf{k}},j} \hbar( v^*_{\textbf{k}}(\textbf{r}_j) \hat{\sigma}_j^\dagger\hat{q}_{\textbf{k}}
 +h.c.),
 \label{ham0}
\end{eqnarray}
where $\omega_{\textbf{k}}$ is the frequency of the ${\textbf{k}}$-th electromagnetic field mode described by the bosonic creation and annihilation operators $\hat{q}_{\textbf{k}}^\dagger$ and $\hat{q}_{\textbf{k}}$; $\nu$ is atomic transition frequency; $\textbf{r}_j$ is the position of $j$-atom and $v_{\textbf{k}}(\textbf{r}_j)$ is interaction coefficient. These constants are defined as $v_{\textbf{k}}(\textbf{r}_j)=-{\textbf{d}}\cdot {\textbf{E}}_\textbf{k}(\textbf{r}_j)$, where $\textbf{d}$ is the atomic dipole moment, and ${\textbf E}_{\textbf{k}}(\textbf{r}_j)$ is the modal field at the atomic position. In the flat homogeneous vacuum eigenmodes are plane waves with wave-vector ${\textbf{k}}$; so, index $\textbf{k}$ in Eq. (\ref{ham0}) denotes both wave-vector and polarisation. Then, the Hamiltonian in the interaction picture rotating with the frequency $\nu$ can be written in the following form
 \begin{equation}\label{Nham}
 \hat{V}(t)=\sum\limits_{\textbf{k},j} \hbar\left( v^*_\textbf{k} \hat{\sigma}_j^\dagger\hat{q}_\textbf{k} e^{-i (\nu-\omega_\textbf{k})t+i\textbf{k}\textbf{r}_j}+\text{h.c.}\right).
\end{equation}
We assume that our system can have no more than one photon. Thus, the general solution for the total system state is described by the
following wave-vector:
\begin{equation}\label{variable}
 \vert \Psi(t) \rangle= \sum\limits_jc_j^{(e,0)}(t)\vert e_j,0\rangle+\sum_{\textbf{k}}c^{(b,\textbf{k})}(t)\vert g,1_{\textbf{k}}\rangle,
\end{equation}
where the state-vector $\vert g,1_{{k}}\rangle$ describes the ground state of all the atoms and a single photon in the $k$-th mode of the field;  $\vert e_j,0 \rangle$ denotes the existed state of $j$-th atom and the vacuum of all the field modes.

Now let us consider a preparation of an excited state of our atomic system by absorption of a photon. We assume this photon to be carried by the plane wave with the wave-vector $\textbf{k}_0$, i.e. 
the initial state of the total system is
\begin{equation}\label{initial_condition}
 \vert \Psi(0)\rangle = \vert g,1_{\textbf{k}_0}\rangle.
\end{equation}

Assuming the weak coupling, we can write the evolution operator of the system as
\begin{equation}\label{evolution_operator}
 U(\tau)=\underleftarrow{T}exp\left\{ -\frac{i}{\hbar}\int\limits_0^\tau dt' \hat{V}(t')\right\}\simeq1-\frac{i}{\hbar}\int\limits_0^\tau dt' \hat{V}(t'),
\end{equation}
where $\underleftarrow{T}$ is the time ordering operator. Then, assuming $v_\textbf{k}\tau \ll 1$ for the time of photon flight through the atomic cloud $\tau$ and  that the incident radiation has been chosen to be
resonant with the atom, i.e., $\nu=\omega_{\textbf{k}}$, the state of the system conditioned on the absorption of the photon is
 \begin{equation}\label{timeddicke}
 \vert \Psi \rangle_{Dicke}\approx \frac{1}{\sqrt{N}}\sum\limits_j e^{i\textbf{k}_0\textbf{r}_j}\vert e_j,0\rangle.
\end{equation}
The entangled delocalized single-excitation (\ref{timeddicke}) is different from the conventional Dicke state by phase factors corresponding to the different phase of the plane wave at the position of each atom. So, for this reason the state (\ref{timeddicke}) was termed "the timed Dicke state" \cite{scully2006}.

Now let us consider the way the state (\ref{timeddicke}) spontaneously decays into the reservoir of the electromagnetic field modes. We assume that the distances between atoms are much larger than the resonant wavelength, and spontaneous decay of each atom occurs independently. Then, implying the Markovian approximation, one can derive the standard equation describing the wave-function of the atom-field system for $j$-th initially completely excited atom decaying into the vacuum of the reservoir, and get that the amplitude of having atom in the excited state decays as $\exp\{-\Gamma t\}$, where $\Gamma$ is the spontaneous decay rate \cite{scullybook}. For times far exceeding the decay rate $\Gamma^{-1}$, the field disentangles from the atom, and the asymptotic field state for the initial state (\ref{timeddicke}) is
\begin{equation}\label{fieldstate}
 \vert \Psi(\infty) \rangle= \frac{1}{\sqrt{N}}\sum_{j,\textbf{k}}\frac{v_\textbf{k} e^{i(\textbf{k}_0-\textbf{k})\textbf{r}_j}}{(\omega_\textbf{k}-\nu)+i\frac{\Gamma}{2}}\vert g,1_{\textbf{k}}\rangle.
\end{equation}
For sufficiently large $N$ the sum of random phases in Eq.(\ref{fieldstate}) gives the Dirac delta-function,
\[\sum_{j}{e^{i(\textbf{k}_0-\textbf{k})\textbf{r}_j}}\propto \delta(\textbf{k}_0-\textbf{k}).
\]

So, a surprising and counter-intuitive effect is arising.  A random set of atoms excited by the plane wave travelling in a certain direction would emit the photon exactly in the propagation direction on the impinging wave. This effect is a very bright manifestation of the role of spatial correlations in the atomic ensemble and was extensively discussed and studied as such.

We aim to consider an influence of the weak gravitation on this effect. Since gravity introduces  inhomogeneity in space, it is natural to expect a deviation from the simple input-output relation discussed above, say, rotation of the emitted photon wave-vector. However, the effect of gravity goes beyond that. It leads to appearance of continuous distribution of directions instead of the delta-function.

\section{Atom-field interaction in the presence of gravitation}

Here we consider how the simple interaction considered in the previous Section is modified in the presence of a weak gravitational field. We will adopt the following approach. Firstly, we consider weak gravitational field as a perturbation for the Maxwell equation in a flat space and find the perturbed eigenmodes. Then, we write down the atom-field interaction Hamiltonian using these new eigenmodes.

\subsection{The metric}

We assume the standard Schwarzschild metric
\begin{eqnarray}
\nonumber
  ds^2=\left(1-\frac{r_{s}}{r}\right)c^{2}dt^{2}-\left(1-\frac{r_{s}}{r}\right)^{-1}dr^{2}-\\
  -r^{2}\left(d\theta ^{2}+\sin ^{2}\theta d\varphi ^{2}\right),
  \label{metrix}
\end{eqnarray}
where $c$ is the speed of light, $t$ is the time coordinate (it can be measured by a stationary clock located infinitely far from the  massive body), $r$ is the radial coordinate, $\theta$ is the colatitude (angle from north pole of the sphere surrounding the  massive body), $\varphi$ is the longitude, and  $r_s$ is the Schwarzschild radius of the massive body, a scale factor which is related to its mass $M$ by $r_s = 2GM/c^2$, where $G$ is the gravitational constant \cite{landau1973theory}.

We consider distances much larger than the Schwarzschild radius, $r \gg r_s$ (for example, for our Earth it is $r_s\simeq1$~cm), and assume that typical size of the considered systems is much less then the gravitating body size, so one can safely neglect tidal effects. Using the coordinate system with axis Z directed along radial coordinate, we take that space don't change along X and Y laboratory axis. So, our metric transforms into:
\begin{eqnarray}\label{korot}
 &&ds^2=h(z)c^2 dt^2-(dx^2+dy^2+1/h(z)dz^2),
\end{eqnarray}
here $h(z)=1-{r_s}/z$, $z$ is the distance from the centre of the massive body.  Then we expand $1/h(z)$ to the first order with respect to $r_s/z_0$ and assume that we are in the close vicinity of the plane  $z=z_0$. After the obvious coordinate change to get manifestly Minkowskian metric at the $z_0$ plane, the final linearized laboratory metric can be written with the help of a constant $a=2 g/c^2$ in the following way:
\begin{eqnarray}
\nonumber
 ds^2=\left(1+a(z-z_0)\right)c^2 dt^2-\\
 (dx^2+dy^2+\left(1-a(z-z_0)\right)dz^2).
 \label{weak field metrics 0}
\end{eqnarray}
 where the acceleration of the free fall in laboratory is $g=GM/z_0^2$. Eq. (\ref{weak field metrics 0}) is the weak field approximation of the Schwarzschild metric which we use in the subsequent discussion. 

\subsection{The perturbed Maxwell equations}

Here we demonstrate an effect of the space curvature on solutions of the Maxwell equations. For a diagonal metrics $g_{\mu\mu}$, $\mu=\overline{0,3}$, one can write the wave equation for the covariant components of the electric field, $E_j$, in the following form \cite{cabral}:
\begin{multline}
g^{jj}\left(\frac{g^{00}}{c^2}\partial_{tt}E_j +\partial^{k}\partial_kE_j\right)=\\
\label{field1}
=a^{kjj}\partial_kE_j+b^{kjj}\partial_jE_k-\\
-g^{jj}(\partial_jg^{kk})\partial_kE_k+g^{jj}\partial_j(g^{00}a^{k00})E_k,
\end{multline}
where $i,k=1,2,3$, the contraction is only on the index $k$, and the coefficients are given by
\begin{align}
\nonumber
a^{k\mu\mu}&=-\partial_k(g^{kk}g^{\mu\mu})-\frac{g^{kk}g^{\mu\mu}}{\sqrt{-g}}\partial_k(\sqrt{-g}), \\
b^{kjj}&=g^{ii}g_{00}a^{k00}-a^{kjj},
\label{coef1}
\end{align}
where $g$ is the determinant of the metric.

To see the effect of the curvature, let us rewrite Eqs. (\ref{field1}) using the approximate Eq. \eqref{weak field metrics 0},
%linearizing coefficients (\ref{coef1}) with respect to $\Delta z = z - z_0$ and $a$, 
obtaining
\begin{widetext}
\begin{eqnarray}\label{maxwell_equations}
\nonumber
(1-a(z-z_0))\frac{\partial_{tt}}{c^2}E_x-\partial_{xx}E_x-\partial_{yy}E_x-(1+a(z-z_0))\partial_{zz}E_x\approx a\partial_zE_x-a\partial_xE_z,\\
\label{lineq}
(1-a(z-z_0))\frac{\partial_{tt}}{c^2}E_y-\partial_{xx}E_y-\partial_{yy}E_y-(1+a(z-z_0))\partial_{zz}E_y\approx a\partial_zE_y-a\partial_yE_z,\\
\nonumber
\frac{\partial_{tt}}{c^2}E_z-(1+a(z-z_0))(\partial_{xx}E_z+\partial_{yy}E_z)-(1+2a(z-z_0))\partial_{zz}E_z\approx a\partial_zE_z.
\end{eqnarray}
\end{widetext}
The system (\ref{lineq}) shows that the curvature of the space leads to coupling between polarisations \cite{cabral}. Also, the field changes with the distance from the gravitation source. So, emitters will interfere differently from the case of the flat space emission of timed Dicke states. Also, one should expect changes in the density of photonic states and the decay rates will become position dependent.  

These simple intuitive considerations are supported by the solution of Maxwell's equations for our linearized metric. 
%We seek the solution in the following form (see Appendix) 
The solution of Maxwell's equations in flat space is represented by 
\begin{equation}\label{electric_field}
 E_j=\sum\limits_\textbf{k}\sum\limits_{s=1}^{2} \alpha_{\textbf{k}}\left[{q}^*_{\textbf{k}s}f_j(\textbf{k},s)e^{i \Theta}+\text{h.c.}\right],
\end{equation}
where $\alpha_{\textbf{k}}$ is amplitude, $\hat{q}_{\textbf{k}s}$ is complex amplitude, 
$f_j(\textbf{k},s)$ is 3-vector of polarisation, $k^\mu=\{c\sqrt{k_x^2+k_y^2+k_z^2},\textbf{k}\}$ being 4-wave vector with three-space part $\textbf{k}=\{k_x,k_y,k_z\}$, $\Theta$ is phase of eigenmodes of electric field.

We seek the solution in curved space as a correction to the flat-space solution linear on the parameter $a$. It is found that for a given $\textbf{k}^{(0)}=\{k_x,k_y,k_z\}$ (see the Appendix):
\begin{widetext}
\begin{eqnarray}\label{amplitude_alpha}
 &&\alpha_{\textbf{k}}\approx\left(\frac{\hbar \omega_\textbf{k}}{2\varepsilon_0 V_0}\right)^{1/2}\left(1+a(z-z_0)\frac{k_x^2+k_y^2}{4k_z^2}\right),\\
 &&\Theta\approx c\sqrt{k_x^2+k_y^2+k_z^2}t-k_x x-k_y y-k_z z  +a \frac{k_x^2+k_y^2+2k_z^2}{4 k_z}(z-z_0)^2,\label{theta}\\ \label{wave_vector}
 &&\tilde{k}_\mu\approx\left\{c\sqrt{k_x^2+k_y^2+k_z^2},\quad -k_x,\quad -k_y,\quad -k_z+a \frac{k_x^2+k_y^2+2k_z^2}{2 k_z}(z-z_0)\right\},
\end{eqnarray}
\end{widetext}
and for 3-vector of polarisation one has 
\begin{eqnarray}\label{polarisation_E}\begin{aligned}
f_1(\textbf{k},s)\approx&f_1^{(0)}(\textbf{k},s)+a\frac{z-z_0}{2}\frac{k_x }{k_z}f_3^{(0)}(\textbf{k},s),\\
f_2(\textbf{k},s)\approx&f_2^{(0)}(\textbf{k},s)+a\frac{z-z_0}{2}\frac{k_y }{k_z}f_3^{(0)}(\textbf{k},s),\\
f_3(\textbf{k},s)\approx&f_3^{(0)}(\textbf{k},s),
\end{aligned}\end{eqnarray}
where  $f_j^{(0)}(\textbf{k},s)$ is the 3-vector of polarisation for the flat space.

We also can write magnetic field $H^i$ in the form similar to Eq. \eqref{electric_field}:
\begin{equation}\label{magnetizing_field}
 H^j=\frac{1}{c \mu_0}\sum\limits_\textbf{k}\sum\limits_{s=1}^{2} \alpha_{\textbf{k}}\left[{q}^*_{\textbf{k}s}p^j(\textbf{k},s)e^{i \Theta}+\text{h.c.}\right],
\end{equation}
where $p^i(\textbf{k},s)$ are the components of contravariant 3-vector 
\begin{eqnarray}\label{polarisation_H}\begin{aligned}
p^i(\textbf{k},s)=\frac{1}{\sqrt{-\tilde{k}^m\tilde{k}_m}}\epsilon^{ijn}\tilde{k}_j f_n(\textbf{k},s),
\end{aligned}\end{eqnarray}
with  contravariant Levi-Civita tensor $\epsilon^{ijn}$. 

Notice that here we assume $k_z\neq0$. It is possible without much difficulties to consider a particular case $k_z=0$. However, we refrain from doing that for simplicity sake. 

Within our weak gravitation approximation, the electromagnetic wave \eqref{electric_field}, \eqref{magnetizing_field} will not lose transversality:
\begin{equation}\label{transversal_E}
-p^i(\textbf{k},s) f_i(\textbf{k},s) =-k^i f_i(\textbf{k},s)=-p^i(\textbf{k},s) k_i =0,
\end{equation}
as expected for the approximation of geometrical optics. 

\subsection{The field Hamiltonian}

To describe the spontaneous emission process, let us quantize the field in a standard way: by replacing  complex modal amplitudes in Eqs. \eqref{electric_field} and \eqref{magnetizing_field} by bosonic creation and annihilation operators. Then, following Ref. \cite{Gorbatsievich}, the Hamiltonian can be introduced as:
\begin{equation}
  \hat{H}=\int d\mathfrak{f}_\mu \varepsilon^\nu \hat T_{\nu}^\mu=\int\limits_{t=const} d^3 x n_\mu \varepsilon^\nu \hat T_{\nu}^\mu,
  \label{ham1}
\end{equation}
where $\varepsilon^\nu=\{c,0,0,0\}$ is the timelike Killing vector of our space-time \eqref{weak field metrics 0}, $\mathfrak{f}_\mu$ is a hypersurface and $n_\mu=\{1/c,0,0,0\}$ is the orthogonal vector to that hypersurface,
$\hat T^{\mu \nu }=\hat F^{\mu \alpha }\hat F^{\nu }{}_{\alpha }-{\frac {1}{4}}g ^{\mu \nu }\hat F_{\alpha \beta }\hat F^{\alpha \beta }$ is the energy-momentum tensor of electromagnetic field $\hat F^{\mu \alpha }$.
Eq. (\ref{ham1}) leads to the following expression 
\begin{equation}\label{interal_for_hamiltonian}
 \hat{H}=\frac{1}{2}\int d^3x  \Big(a(z-z_0)-1\Big)\left(\varepsilon_0 \hat E_j \hat E^j+\mu_0 \hat H_j \hat H^j\right).
\end{equation}

Also, \eqref{interal_for_hamiltonian} matches to the Hamiltonian from optical analogy \cite{landau1973theory}:
\begin{eqnarray}\nonumber
 \hat{H}=\frac{1}{2}\int d^3 x  \left(-\hat E_j \hat D^j-\hat H_j \hat B^j\right),
\end{eqnarray}
where $D^j=\varepsilon_0 E^j/\sqrt{h(z)}$ is electric displacement field, $B^j=\mu_0 H^j/\sqrt{h(z)}$ is magnetic displacement field. 

To express the Hamiltonian in terms of the optical modes, one needs to carry on summation over modes and polarisation. This procedure is non-trivial only for $z$-coordinate. Noticeably, expansion over the $a$-parameter brings about dependence on the first and the second degree of $z$. However, these terms are mutually cancelling. 

The quantisation volume in curved space acquires dependence on the metric, $g_{\mu\nu}$: 
\begin{eqnarray}
\label{vcurv}
 V^{curv}=\int \sqrt{-\gamma}d^3 x=L^3\left(1-\frac{a}{2}\left(Z-z_0\right)\right),
\end{eqnarray}
where $Z$ is position of centre of the cubic quantisation volume with the edge length $L$ and $\gamma$ is determinant of space part of metric \eqref{korot}.
In the momentum space the element of volume is inversely proportional to the elementary volume in the coordinate space. So, an integration over the momentum space is as follows
\begin{equation}\label{sumk}
 \sum\limits_{\textbf{k}'}\rightarrow \left(1+\frac{a}{2}\left(Z-z_0\right)\right) \frac{V_0}{(2\pi)^3}\int dk_x' dk_y' dk_z',
\end{equation}
for each polarisation vector.

%This takes place due to the fact that \eqref{interal_for_hamiltonian} depends linearly  on $h(z)$ in , and

Discarding the vacuum energy, after some algebra from Eq.~(\ref{interal_for_hamiltonian}) one gets the Hamiltonian in following form:
\begin{equation}\label{hamiltonian}
 \hat{H}=\hbar \sum\limits_\textbf{k} \omega_\textbf{k}[Z-z_0] \hat q_{\textbf{k}}^\dagger \hat q_{\textbf{k}}^{},
\end{equation}
where $\omega_\textbf{k}[z]= \omega_\textbf{k}(1+a z/2)$ (the notation $[z]$ will be also used further  for emphasising $z$-dependence of different variables), $\omega_\textbf{k}=c\sqrt{k_x^2+k_y^2+k_z^2}$. We should emphasise here that the set of modes $\{\textbf k\}$ is determined for the height $Z$, and the coordinate time is a proper time for the height $z_0$.

\section{The single atom decay}

Now let us consider changes in the spontaneous emission of an individual two-level atom due to the presence of gravitational field. For an atom at $z=z_0$ the Hamiltonian in the interaction picture has the following form
\begin{eqnarray}
\label{hmod1}
 \hat{V}(t)=\sum\limits_\textbf{k} \hbar\left( v^*_\textbf{k}(\textbf{r}_{at}) \hat{\sigma}^\dagger\hat{q}_\textbf{k} e^{i (\nu-\omega_\textbf{k}[Z-z_{at}])t}+\text{h.c.}\right),
\end{eqnarray}
where we calculate the interaction Hamiltonian as $-d^j E_j(\textbf r_{at})$. Notice that we assumed the coordinate system centred on the atom reducing all the height corrections. 
Eq.~(\ref{hmod1}) leads to the following  equation for the excited state amplitude \cite{scullybook}:
\begin{eqnarray}\begin{aligned}\label{vk2}
\dot{c}^{(e,0)}(t)= & -\sum\limits_\textbf{k}\vert v_\textbf{k}(\textbf{r}_{at})\vert^2\times  \\ &\times\int dt' e^{i (\nu-\omega_\textbf{k}[Z-z_{at}])(t-t')}c^{(e,0)}(t').
\end{aligned}
\label{crel1}
\end{eqnarray}

The difference from the flat space case manifests itself in changing of the mode quantization volume according to \eqref{sumk}, and the presence of correction to  $\omega_\textbf{k}$. 

After applying in the standard way the Markovian approximation, integrating over $\textbf{k}$ and $dt'$,  and taking the linear approximation on $a$, one gets the following amplitudes of excited atomic state $c^{(e,0)}$, and the modal single-photon amplitudes  $\dot{c}^{(g,\textbf{k})}$: 
 \begin{eqnarray*} \label{solA}
&&c^{(e,0)}=\exp\left[-\frac{\Gamma}{2} t\right],\\ \label{solB}
&& \dot{c}^{(b,\textbf{k})}=-i v_{\textbf{k}}(\textbf{r}_{at})\exp\left[-i(\nu-\omega_\textbf{k}[Z-z_{at}])t-\frac{\Gamma}{2}t\right],
\end{eqnarray*}
where $\Gamma$ is the spontaneous decay rate at the height $z_{at}$. The amplitudes are written for the proper time of atom on height $z_{at}$ and $\Gamma$ is height-independent only in proper time. The changing of reference system from the location of the atom at $z_{at}$ to the arbitrary location of laboratory frame at $z$ changes the time according to change in metric (\ref{weak field metrics 0}) as
\begin{equation}
 t_{at}\longrightarrow t \left(1+\frac{a}{2}(z_{at}-z)\right),
\end{equation}
and the wavevector $\textbf{k}$ at $z_{at}$  to the wavevector $\textbf{k}'$ for $z$.
Asymptotically, the modal single-photon amplitudes in the laboratory frame are
\begin{multline}\label{solB}
 c^{(b,\textbf{k}')}=\frac{v_{\textbf{k}'}(\textbf{r}_{at})}{\omega_{\textbf{k}'}[Z-z]-\nu[z_{at}-z]-\frac{i}{2}\Gamma[z_{at}-z]}=\\
 =\frac{v_{\textbf{k}'}(\textbf{r}_{at})(1-\frac{a}{2}(z_{at}-z))}{\omega_{\textbf{k}'}[Z-z_{at}]-\nu-\frac{i}{2}\Gamma},
\end{multline}
where $\nu[z]=\nu(1+a z/2)$ and $\Gamma[z]=\Gamma(1+a z/2)$. 

As it should be expected, the result coincides with the one obtained by considering the total Hamiltonian in non-local reference system. In that case we should deal with the effect of the red-shift of two-level system \cite{pound1959gravitational, Gorbatsievich} as $\nu[z_{at}-z]$ and consider corrections of electric field \eqref{electric_field} in calculation of $\vert v_k \vert^2$ in \eqref{vk2} that will lead to the position-dependence of the decay rate,  $\Gamma[z_{at}-z]$. These are the well-known effects of the time dilation predicted by the general theory of relativity.

% Thus, we are getting the following solution for the amplitudes of single-photon states
% 
% Eqs. (\ref{solB}) shows the manner of the gravitation dephasing of timed Dicke states: both phases and amplitudes of each single-photon component of the emitted state will not add up as it was for the timed Dicke state in the flat space. 

\section{The timed Dicke state in curved space}

In the considered ensemble, the atoms are situated at different heights. One should take that into account when deriving the general Hamiltonian for the ensemble. Thus, in the usual dipole and the rotating-wave approximations, the Hamiltonian describing interaction of just one field mode $(\textbf{k}_0)$ with the ensemble is
\begin{gather}
     \label{hmod}
    \hat{V}(t)=\text{\hfill{}}\\
    =\hbar\sum\limits_{j} \Big( \tilde{v}^*_{\textbf{k}_0}(\textbf{r}_{at}) \hat{\sigma}^\dagger\hat{q}_{\textbf{k}_0} e^{i (\nu[z_j-z_0]-\omega_{\textbf{k}_0}[Z-z_0])t}+
    \text{h.c.}\Big),
    \nonumber\\\nonumber
    \tilde{v}^*_{\textbf{k}_0}(\textbf{r}_{at})=v^*_{\textbf{k}_0}(\textbf{r}_{at})\Big(1+a F_{\textbf{k}_0}\{z_j-z_0\}\Big),\\
    \nonumber
    F_{\textbf{k}_0}\{z\}=z\beta_{\textbf{k}_0}+iz^2 \gamma_{\textbf{k}_0},
\end{gather}
where $\beta_{\textbf{k}_0}$ and $\gamma_{\textbf{k}_0}$ are real constants depending on $\textbf{k}_0$ and the atomic dipole moment. The dependence on $(z_j-z_0)$ in $\hat{V}(t)$ is stemming from modification of the polarisation vector \eqref{polarisation_E},  amplitude  \eqref{amplitude_alpha} and  phase \eqref{theta} of the modal field.

To find the timed Dicke state in the curved space, we follow the way of deriving it in flat space with the same assumptions (namely, weak coupling and closeness of frequencies of incident radiation and the atomic transition). First, we introduce the evolution operator as \eqref{evolution_operator} using the interaction Hamiltonian \eqref{hmod} and then apply it to the initial state \eqref{variable}. So, we have the following result for the atomic cloud state conditioned on the photon absorption:
% 
% The timed Dicke state in flat space is given by \eqref{timeddicke}. The interaction Hamiltonian in the form (\ref{hmod}) gives the following result for the atomic cloud state conditioned on the photon absorption:
\begin{equation}\label{state_psi}
 \vert \Psi\rangle_{Dicke}^{curv}\approx\frac{1}{\sqrt{N}}\sum\limits_j e^{i\textbf{k}_0\textbf{r}_j}\left(1+a F_{\textbf{k}_0}\{z_j-z_0\} \right)\vert e_j,0\rangle,
\end{equation}
where $\textbf{k}_0$ is the wave-vector of plane wave at height $z_0$ assumed to be a centre of the atomic cloud. 
% The function  $F_{\textbf{k}_0}(z_j)$ is 
% \begin{equation}\label{function_F}
%     F_{\textbf{k}_0}(z_j)=(z_j-z_0)\frac{k_x^2+k_y^2}{4k_z^2}-i(z_j-z_0)^2\frac{k^2+k_z^2}{4k_z},
% \end{equation}
% here the first degree of $z_j$ is appeared from modification of amplitude of electromagnetic wave \eqref{amplitude_alpha} (real) and the second -- from the modification of phase \eqref{theta} (imaginary). Also in expression \eqref{state_psi}
With the time of photon flight through the atomic cloud, $\tau$, being assumed to be small, here we take $v_\textbf k \tau\ll 1$, and also neglect terms proportional to $\tau^2$.

\section{The dephasing}

Now let us see how the position-dependence given by Eq.~\eqref{solB} influence the angular distribution of the emitted field.

So, from integration of the amplitude \eqref{solB}, for $t\rightarrow\infty$ %up to the terms of the second order on the constant $a$,
we can write the state of the emitted photon as
\begin{eqnarray}\label{curved_dicke_state}\begin{aligned}
 \vert \Psi(\infty) \rangle^{curv}\approx\frac{1}{\sqrt{N}}\sum_{j,\textbf{k}}v_{\textbf{k}} e^{i(\textbf{k}_0-\textbf{k})\textbf{r}_j}\times\\ 
 \times\left(\frac{1}{(\omega_\textbf{k}[Z-z_j]-\nu)+i\frac{\Gamma}{2}}+O(a)\right)\vert g,1_{\textbf{k}}\rangle.
%+\\
%   &+\frac{a}{\sqrt{N}}\sum_{j,\textbf{k}}v_{\textbf{k}}
%  \frac{e^{i(\textbf{k}_0-\textbf{k})\textbf{r}_j}}{(\omega_\textbf{k}[Z-z_j]-\nu)+
%  i\frac{\Gamma_0}{2}}{F}_{\textbf{k}_0}[z_j-z_0]\vert g,1_{\textbf{k}}\rangle.
\end{aligned}\end{eqnarray}

Notice that to make more clear the subsequent derivation, in Eq.~(\ref{curved_dicke_state}) we show only terms essential for the final result, i.e. we have already included a part of
terms linear on $a$  (e.g. the function $F_{\textbf{k}_0}\{z-z_0\}$ from Eq. \eqref{state_psi} describing the deviation of the generated timed Dicke state from its flat-space analogy) into $O(a)$, while keeping others. 
%assuming them to be much less than unity (including  the function $F_{\textbf{k}_0}[z-z_0]$ from Eq. \eqref{state_psi} describing the deviation of the generated timed Dicke state from its flat-space analogy) including them in the terms $O(a)$. However, for the moment we do not omit the similar terms stemming from the fraction in Eq.(\ref{curved_dicke_state}). 

 %Explicit form of to be not important for the effect of interest and will not be presented here.
We consider a large number of atoms  homogeneously distributed in the large volume $V_0$. Also, we assume validity of the typical assumptions for geometric optic approximation: the spatial extension of the atomic ensemble is much more than wavelength, but much less than radius of curvature of our space-time. Thus, we can replace the summation over the positions with integration as
\begin{equation*}
 \sum_j\longrightarrow \frac{N}{V_0}\int d^3 x \sqrt{-\gamma},
\end{equation*}
take only the leading term in $a$ and obtain the following expression for the field wave-function:
\begin{multline}\label{integral}
 \vert \Psi(\infty) \rangle^{curv}\approx\frac{\sqrt{N}}{V_0}(2\pi)^2\sum_{\textbf{k}}v_\textbf{k}\delta({k_{0}}_x-k_x)\delta({k_{0}}_y-k_y)\times
 \\
\times \Bigg(\int dz 
 \frac{ e^{i({k_{0}}_z-k_z)z}}{(\omega_\textbf{k}-\nu+i\frac{\Gamma}{2})+\frac{a}{2}\omega_\textbf{k}(Z-z)}+O(a)\Bigg)\vert g,1_{{k}}\rangle.
\end{multline}
%\begin{color}{blue}
%\begin{multline}\label{integral}
% \vert \Psi(\infty) \rangle^{curv}\approx\frac{\sqrt{N}}{V_0}(2\pi)^2\sum_{\textbf{k}}v_\textbf{k}\delta({k_{0}}_x-k_x)\delta({k_{0}}_y-k_y)\times
% \\
%\times \Bigg(\int dz 
% \frac{ e^{i({k_{0}}_z-k_z)z}}{(\omega_\textbf{k}-\nu+i\frac{\Gamma}{2})+\frac{a}{2}\omeg%a_\textbf{k}(Z-z)}+\\+\int dz  e^{i({k_{0}}_z-k_z)z} O(a)\Bigg)\vert g,1_{{k}}\rangle.
%\end{multline}
As mentioned, only the essential first integral inside the brackets of Eq. \eqref{integral} gives non-trivial deviation from the flat-space solution.
%Notice that up to our assumed level of accuracy, the functions $F_{\textbf{k}_0}[z-z_0]$ describing the deviation of the generated timed Dicke state from its flat-space analogy, does not contribute to the leading term in $a$ of $\vert \Psi(\infty) \rangle^{curv}$. 
%Up to the terms of the second order on the constant $a$

As it should be expected, the components of the wave-vector along $x$ and $y$ axes do not change. 
Up to a phase factor, Eq.~\eqref{curved_dicke_state} leads to
\begin{eqnarray}\nonumber
\begin{aligned}
 &\vert \Psi(\infty) \rangle^{curv}\sim\frac{\sqrt{N}}{V_0}(2\pi)^3\times\\ &\times\sum_{k_x,k_y,k_z}v_\textbf{k}\delta({k_{0}}_x-k_x)\delta({k_{0}}_y-k_y)
  \Big(G_{k_z}+O(a)\Big)\vert g,1_{{k}}\rangle.
\end{aligned}\end{eqnarray}
%\begin{eqnarray}\nonumber
%\begin{aligned}
% &\vert \Psi(\infty) \rangle^{curv}\sim\frac{\sqrt{N}}{V_0}(2\pi)^3\sum_{k_x,k_y,k_z}\delt%a({k_{0}}_x-k_x)\delta({k_{0}}_y-k_y) \times \\
% &\times 
%  v_\textbf{k}\Big(G_{k_z}\big(1+O(a)\big)+\delta(k_{z0}-k_z)O(a)\Big)\vert %g,1_{{k}}\rangle.
%\end{aligned}\end{eqnarray}
where  for $\Gamma\ll\nu$ and using $\omega_{\textbf{k}_0}=\nu$ we have
\begin{eqnarray}
G_{k_z}=\frac{-i}{a\nu}\exp\left[-({k_{0}}_z-k_{z})
 \frac{\Gamma}{a\nu}\right] \Theta [{k_{0}}_z-k_{z}],
\label{g}  
\end{eqnarray}
where $\Theta [{k_{0}}_z-k_{z}]$ is the Heaviside step function.

Eq. (\ref{g}) gives us the expected coincidence of propagation directions for impinging  and 
emitted photons for flat space, $\lim\limits_{a\rightarrow 0}G_{k_z}\propto\delta({k_{0}}_z-k_z)$.
 For the non-zero curvature ($a\neq 0$), we have a finite width distribution of directions around the vector ${k_{0}}_z$. Thus, the spread of wave-vectors around the wave-vector of the impinging photon is defined by the quantity ${a \nu}/{\Gamma}\cos\theta_0$, %probably cosine is excessive
here $\theta_0$ is the angle between $z$-direction and $\textbf{k}_0$. Notice that this spread is asymmetric, the photon tends to deviate toward the direction of gravitational attraction. Also, not only the emission direction is "blurred". The photon energy is changed, so the external observer will see the superposition of photon wave-packets with frequencies different from the original one, with the width of this frequency spread being 
\begin{equation}
 \delta\omega\approx\frac{a c \nu}{\Gamma}\cos\theta_0.
 \label{deff}
\end{equation}
Eq. \eqref{deff} demonstrates that indeed the gravitation influence quantum interference in the process of spontaneous emission of a delocalized collective single-excitation state. As the result, for the emitted field this influence looks like dephasing. Of course, it is not large. Near the Earth surface $a=2\cdot10^{-16}$~1/m, for typical values of the spontaneous emission rate $\Gamma\approx10^8$~Hz and optical frequencies $\nu\approx 10^{15}$~Hz, the frequency spread \eqref{deff} is about several Hz. But near the stronger gravitating bodies (especially in the vicinity of black holes) this effect will be considerably more pronounced. 

Thus, even classical gravitation do affect the quantum interference in a destructive way. 

\section*{Conclusions}

We have considered here the process of creation of the timed Dicke states in the presence of weak gravitational field, and emission of these states. In difference with the flat space, the ensemble of randomly placed two-level atoms does not emit the photon in the direction of the photon that was absorbed. Gravitation leads to appearance of an asymmetric distribution of directions. The photon is not unexpectedly deviating toward the gravitational attraction. But also the photon frequency appears to by changed. Instead of just one plane wave, one has a superposition of single-photon wave-packets with different frequencies. 

Of course, the model considered here (just like the original model of timed Dicke state emission for the flat space \cite{scully2006}) can hardly be realised in practice without accounting for example, for atomic movement, recoil, finite temperature effects, etc. (however, one can even now pinpoint some good candidates for realising the timed Dicke states, for example, colour centres in diamonds \cite{diamonds}). Nevertheless, the demonstrated effects of interplay between quantum and classical interference in the presence of the gravitation show very clearly how the influence, considered extremely weak at once, may in fact noticeably change the outcome of an intrinsically quantum process. We have shown that even classical gravitation does affects  quantum interference producing something close to the decoherence effects so ubiquitous in the quantum world.

\acknowledgments

The authors gratefully acknowledge financial support from the Belarus state scientific program "Convergence-2020".

\appendix
\section{Solution of Maxwell equations}
The Maxwell equations linearized on the parameter $a$ are given by expressions \eqref{maxwell_equations}. We will look for the solution in the form of a perturbed plane wave with the perturbation proportional to $a$. 
For covariant vector of negative part of electric field eigenmode we have:
 \begin{equation*}
  E^{(-)}_{j}(\textbf{k},s)\sim e^{i \Theta^{(0)}}\alpha_{\textbf{k}}^{(0)} f_j^{(0)}(\textbf{k},s)\left(1+a M_j(z)\right){q}_{\textbf{k},s}^{*},
 \end{equation*} 
where $ \Theta^{(0)}=c \sqrt{k_x^2+k_y^2+k_z^2}t-k_x x-k_y y-k_z z$ is phase,
$\alpha_{\textbf{k}}^{(0)}$ is amplitude, $f_j^{(0)}$ is $j$-component of vector of polarisation of a flat plane wave. The function $M_j(z)$ represents a linear perturbation to the plane wave and consists of real and imaginary parts for amplitude/polarisation and phase modifications respectively. It is described by the following equations
\begin{widetext}\begin{eqnarray}
\nonumber
&(k_x^2 +k_y^2+2k_z^2) (z-z_0)+2 i k_z M_3'(z)+i k_z-M_3''(z)=0,\\
\nonumber&f^{(0)}_1(\textbf{k},s)\left((k_x^2 +k_y^2+2k_z^2) (z-z_0)+2 i k_z M_1'(z)+i k_z-M_1''(z)\right)=i f^{(0)}_3 (\textbf{k},s)k_x,\\
\nonumber&f^{(0)}_2(\textbf{k},s)\left((k_x^2 +k_y^2+2k_z^2) (z-z_0)+2 i k_z M_2'(z)+i k_z-M_2''(z)\right)=i f^{(0)}_3(\textbf{k},s)k_y,
\end{eqnarray}
with the solutions
\begin{eqnarray}
\label{A2}
 M_3(z)&=&(z-z_0)\frac{k_x^2+k_y^2}{4 k_z^2}+
 i(z-z_0)^2\frac{k_x^2+k_y^2+2k_z^2}{4 k_z}+C_3,\\
M_1(z)&=&(z-z_0)\frac{k_x^2+k_y^2}{4 k_z^2}+i(z-z_0)^2\frac{k_x^2+k_y^2+2k_z^2}{4 k_z}+\frac{k_x}{2k_z}(z-z_0)\frac{f^{(0)}_3(\textbf{k},s)}{f^{(0)}_1(\textbf{k},s)}+C_1,\label{A3}\\
M_2(z)&=&(z-z_0)\frac{k_x^2+k_y^2}{4 k_z^2}+i(z-z_0)^2\frac{k_x^2+k_y^2+2k_z^2}{4 k_z}+\frac{k_y}{2k_z}(z-z_0)\frac{f^{(0)}_3(\textbf{k},s)}{f^{(0)}_2(\textbf{k},s)}+C_2,\label{A4}
\end{eqnarray}
\end{widetext}
where  $C_j$ is some complex constants (the first integrating constant) and we ignore term $\propto exp\{2ik_z z\}$ (the second integrating constant is supposed be equal to zero). The same real part in \eqref{A2}-\eqref{A4}  interprets as amplitude modification \eqref{amplitude_alpha}, the same imaginary -- as phase modification \eqref{theta}, all that's left is modification of polarisation vector. To determine constants $C_j$ we use the Gauss' law \cite{cabral}, that in  metric \eqref{weak field metrics 0} looks like
\begin{equation}
 \partial_x E_x+\partial_y E_y+\left(1+a(z-z_0)\right)\partial_z E_z=0.
\end{equation}
Its solution allows us to take $C_1=C_2=0$, and $C_3=-i(k_x^2+k_y^2)/(4k_z^3)$.
Note that all modifications are obtained here in approximation linear in $a$. They can be found also directly from geometrical optics approximation of Maxwell equations \cite{MTW}, except for the component of $f_3 (\textbf{k},s)$, because it has the constant term, proportional to wavelength that belongs to post-geometrical approximation. We will neglect it, because the accounting of only linearized metric cannot be enough to work with post-geometrical terms. So, the modification of polarisation vector is given by Eq. \eqref{polarisation_E}.

The contravariant vector of magnetic field $H^j$ of an eigensolution can be found in the form similar to the electric field:
$ {H^j}^{(-)}_{\textbf{k},s}\sim e^{i \Theta}\alpha_{\textbf{k}} p^j(\textbf{k},s){q}^*_{\textbf{k}s}$,
with modified $\alpha_{\textbf{k}}$ of \eqref{amplitude_alpha} and $\Theta$ of \eqref{theta}. Maxwell equations for $H^j$ are
\begin{equation*}
 \left(1-a(z-z_0)\right) \partial_t H^j=-e^{jlm}\partial_l E_m,  
\end{equation*}
where $e^{jlm}$ is Levi-Civita symbol. For geometrical optics approximation wave vector can be found as $\tilde{k}_\mu=\partial_\mu \Theta$ \eqref{wave_vector}. Using the explicit form of $\Theta$ we can find contravariant components of $p^j(\textbf{k},s)$:
\begin{widetext}
\begin{eqnarray}\label{r_from_H}\begin{aligned}
p^1(\textbf{k},s)&=-\frac{\tilde{k}_2 f_3(\textbf{k},s)(1-i a\frac{1}{2k_z})-\tilde{k}_3f_2(\textbf{k},s)(1+i a\frac{k_x^2+k_y^2}{4 k_z^3})}{\sqrt{k_x^2+k_y^2+k_z^2}}(1+a(z-z_0)),&\\
p^2(\textbf{k},s)&=-\frac{-\tilde{k}_1 f_3(\textbf{k},s)(1-i a\frac{1}{2k_z})+\tilde{k}_3f_1(\textbf{k},s)(1+i a\frac{k_x^2+k_y^2}{4 k_z^3})}{\sqrt{k_x^2+k_y^2+k_z^2}}(1+a(z-z_0)),&\\
p^3(\textbf{k},s)&=-\frac{-\tilde{k}_2 f_1(\textbf{k},s)+\tilde{k}_1f_2(\textbf{k},s)}{\sqrt{k_x^2+k_y^2+k_z^2}}(1+a(z-z_0)).&
\end{aligned}\end{eqnarray}
\end{widetext}
If we neglect post-geometrical terms in \eqref{r_from_H}, the $p^j(\textbf{k},s)$ can be written in the form \eqref{polarisation_H}:
\begin{eqnarray}\begin{aligned}
p^j(\textbf{k},s)=\frac{1}{\sqrt{-\tilde{k}^m\tilde{k}_m}}\epsilon^{jlk}\tilde{k}_l f_k(\textbf{k},s),
\end{aligned}\end{eqnarray}
with Levi-Civita tensor $\epsilon^{jlk}=-\sqrt{h(z)}e^{jlk}$ ($\epsilon_{jlk}=1/\sqrt{h(z)}e_{jlk}$) and $\sqrt{-\tilde{k}^m\tilde{k}_m}=\sqrt{k_x^2+k_y^2+k_z^2}/\sqrt{h(z)}$.

\end{document}